\documentclass[a4paper,12pt]{article}
\usepackage{amssymb, amsmath,bbold}
\usepackage{graphicx}
\usepackage{overpic}

\def\balpha{\boldsymbol{\alpha}}

\title{\begin{large}\textbf{The Rotating Mass Matrix, the Strong CP Problem and Higgs Decay}\end{large}}
\author{Michael J BAKER\thanks{bakerm@maths.ox.ac.uk}\and TSOU Sheung Tsun\thanks{tsou@maths.ox.ac.uk}}
\date{\begin{normalsize}Mathematical Institute, University of Oxford,\\Oxford, OX1 3LB, United Kingdom\end{normalsize}}
\begin{document}

\maketitle{}

\begin{abstract}
\noindent
We first show that the rotating mass matrix hypothesis
suggested earlier, where the massive
eigenvector of a rank-one mass matrix
changes with renormalization scale, is consistent
with the latest experimental data on fermion mass hierarchy and
mixing, including the CP
violating KM phase.  
We obtain thereby a smooth trajectory for the massive
eigenvector as a function of the scale.
Using this trajectory we next study Higgs decay and find suppression of 
$\Gamma(H\rightarrow c\bar{c})$ compared to the standard model
predictions for a range of Higgs masses.  We also give limits for flavour
violating decays, including a relatively large branching ratio for the 
$\tau^-\mu^+$ mode.
\end{abstract}

\section{Introduction}
Even though the standard model is a tremendously successful theory,
there are many aspects of it which are not understood.  Among the most
puzzling of these is the hierarchy of masses across generations.  The
rotating mass matrix picture 
purports to give a natural answer to the mass hierarchy problem.  
It also relates fermion masses to the CKM and MNS matrices, an
appealing 
feature which has previously been considered in other contexts, 
e.g., \cite{Fritzsch:1979zq}.

By a rotating mass matrix we mean a rank-one mass matrix whose top
eigenvector changes direction in generation space as
a result of its evolution with renormalization scale.  In this paper
we wish to show that the rotating mass matrix hypothesis is 
fully consistent
with the current experimental data on both the fermion masses and 
mixing angles,
except for the masses of the lightest quarks
for which the implication of the hypothesis is not clear.  
This work represents a
serious advance from previous tests, which were either model-dependent 
\cite{Bordes:1997ft}, or limited to 2 generations by available data at
the time \cite{Bordes:2002nh}; and none of those could incorporate the
CP phase in the CKM matrix as we now can, using a newly suggested 
mechanism \cite{Bordes:2010nb}.

We find a smooth trajectory for the top eigenvector
which results in a very good CKM matrix,
with all the relevant parameters within experimental error, giving a
strong CP-violating theta-angle of order unity.  We then apply these
results to Higgs decay and some
flavour-violating decays for a range of Higgs masses.
The Higgs decay
presents interesting and testable features consistent with the
predictions of \cite{Bordes:2009wi}.
The flavour-violating
decays are found to be all within experimental bounds.

\section{Consistency of the Rotating Mass Matrix Hypothesis 
with Data}
\subsection{Preliminaries}

The use of a rotating mass matrix to generate lower generation masses
and nontrivial mixing was suggested in \cite{Bordes:1997ft} and further
described in  \cite{Chan:2006nf}.  Here we need only to very briefly 
recall its essential features, and introduce some notations and
formulae to be used in the phenomenological fit.

Following Weinberg \cite{Weinberg:1973ua} we put the mass matrix in a
hermitian form without $\gamma_5$.  If, in addition, it is rank-one, then we can write
\begin{equation}
\label{rank_one_mass_matrix}
m=m_T \balpha \balpha^\dagger
\end{equation}
where $m_T$ is the non-vanishing eigenvalue of $m$ and $\balpha$ is
its normalised eigenvector.  We shall assume that only $\balpha$,
identical for the up- and down-type quarks,
changes under renormalization and not the eigenvalue $m_T$, which depends
only on the type of fermions.  In this work we
are concerned only with a phenomenological fit and so do not 
refer to any specific renomalization scheme.  Models for this were
treated in \cite{Bordes:1997ft, Chan:2006nf}.  
For ease of
presentation we shall now give formulae for the up-type quarks, so that
$m_T=m_U$.  The cases for the down-type quarks and leptons are similar.

  In renormalization theory it is usual to define particle masses at a 
scale equal to the mass itself.  Similarly, in the current scheme we define 
the state vector of the top quark, $\mathbf{v}_{t}$, as the massive 
eigenvector at the scale of the top quark, $\mathbf{v}_{t} =
\balpha_{t}$, 
where we denote $\balpha(\mu=m_{i})$ by $\balpha_{i}$.  The eigenvalue
of the mass matrix at this scale will then give the top quark mass, so
$m_U=m_t$.  We then run the scale to the mass of the charm quark.  The
particle state vectors $\mathbf{v}_t$ and $\mathbf{v}_c$ must be
orthogonal,  but the eigenvector $\balpha_{c}$ need not be orthogonal 
to $\mathbf{v}_{t}$.  We thus take the charm quark state vector to lie 
in the subspace orthogonal to $\mathbf{v}_{t}$ in the direction of
$\balpha_{c}$, i.e., $\mathbf{v}_{c}\ ||\
\textrm{proj}_{\mathbf{v}_{t}^{\bot}}(\balpha_{c})$.  At $\mu = m_{c}$
the charm quark state vector, $\mathbf{v}_{c}$, has in general
a component in the
direction of the massive eigenvector $\balpha_{c}$.  
Thus the particle state acquires some mass via a `leakage' from the 
massive state at $\mu=m_{c}$, even though it was massless at $\mu =
m_{t}$.  One can think of an effective $\balpha^{\textrm{eff}} \in 
\mathbf{v}^{\bot}_{t}$, $\balpha^{\textrm{eff}}= 
\textrm{proj}_{\mathbf{v}_{t}^{\bot}}(\balpha)$.  This 
$\balpha^{\textrm{eff}}$ will correspond to the eigenvector of a 
$2 \times 2$ rank-one matrix (since at this scale only the up and
charm quarks can be produced), whose eigenvalue at $\mu=m_c$ will be 
$m_c$.  We then follow a similar procedure to define $\mathbf{v}_{u}$.  
This procedure gives the masses of the up-type quarks to be
\begin{eqnarray}
\label{masses}
m_t & = & m_U,\notag\\
m_c & = & m_U (|\balpha_c\cdot\mathbf{v}_c|)^2,\notag\\
m_u & = &  m_U (|\balpha_u\cdot\mathbf{v}_u|)^2.
\end{eqnarray}

Thus all the quark states get nonzero masses, as experimentally
indicated, while the mass matrix itself remains rank-one with two
zero eigenvalues at any scale.   This important property of the
rotating mass matrix, which may at first sight be surprising, comes
about because of unitarity which requires that the mass matrix be
truncated to remove those states below the threshold at which these
states cease to be physical states.  
Hence in this `leakage' mechanism physical masses are
eigenvalues of the physical (truncated) mass matrix and not
necessarily eigenvalues of the rotating mass matrix, and so can be
nonzero 
while the rotating mass matrix retains its two zero eigenvalues at all
scales.
The fact that we should truncate the mass matrix at scale
thresholds (as at $\mu=m_c$ above) is similar to what one has to do
to the multi-channel $S$-matrix \cite{lecouteur, newton}.  At scales
below the highest channel the $S$-matrix will have to
be truncated from say $n
\times n$ to $(n-1) \times (n-1)$ in order to maintain unitarity,
while the $S$-matrix itself remains analytic at all scales.  For
further discussions on these points see, e.g., \cite{Bordes:2007vu}.

It should be noted that due to confinement in the strong sector the 
relation between experimental measurements of the light quark masses
and the definition we adopt above is unclear.  For this reason we do
not include the masses of the $u$ and $d$ quarks in this study. 

As the rotation of $\balpha$ will be driven by some renormalization 
group equation we shall be interested in finding a smooth trajectory.
Although in specific models
 \cite{Bordes:1997ft, Chan:2006nf} so far constructed
all particle types share the same trajectory, this is not necessary 
in general.  We find, however, that we can well accommodate all 
particle types on the same trajectory.  

The state vectors of both the up- and down-type triads are
determined at scales equal to the masses of the second generation, so that the above
procedure determines all six quark state vectors.  This in turn gives
a matrix $V$ 
\begin{equation}
\label{up_down}
V_{ij}=\langle\textbf{v}^{\textrm{up}}_{i} | \textbf{v}^{\textrm{down}}_{j}\rangle
\end{equation}
where $| \textbf{v}^{\textrm{up}(\textrm{down})}_{i} \rangle$ are the
state vectors so obtained.   This matrix $V$
would be the CKM matrix if there was no extra
contribution from the theta-angle term in the QCD Lagrangian, 
the effects of which will now have to be considered.

It has long been known \cite {Weinbergbook}
that if there is at least one massless quark then the CP violating
term in the QCD Lagrangian
can be absorbed by a chiral transformation on the massless quark field.
The quark must be massless since otherwise this chiral transformation would make its mass parameter complex which would in turn lead to CP violating effects which are not seen \cite{Baker:2006ts}.

Now the rotating mass matrix scheme has, as noted above, the
special property that the matrix can remain rank-one, i.e.,
with two zero eigenvalues, while giving nonzero masses to all the quarks.
This means that one can rotate away any theta-angle term in
the QCD action by a chiral transformation without the necessity of
having a physical massless quark \cite{Bordes:2007vu}.  Further, 
as is shown in \cite{Bordes:2010nb}, the phase
removed by the chiral transformation gets transmitted by the rotation
to the CKM matrix and gives rise there to a Kobayashi-Makawa
CP-violating phase, even if the matrix in \eqref{up_down} is real.  
If we set up a Darboux triad consisting of the 
radial vector $\balpha(\mu)$, the
tangent vector $\boldsymbol{\tau}(\mu)\ ||\ \dot{\balpha}(\mu)$ and 
the normal vector $\boldsymbol{\nu}(\mu)$ orthogonal to both, then it
is argued in
\cite{Bordes:2010nb} that in order to preserve hermiticity of the mass
matrix the chiral transformation should be effected in the normal direction
$\boldsymbol{\nu}(\mu)$.
  If we
choose axes such that $\boldsymbol{\nu}$ is the third axis,
then the chosen chiral transformation will give rise to a phase rotation
of the left-handed fields

\begin{equation}
P_0=\left(\begin{array}{ccc}
1&0&0\\
0&1&0\\
0&0&e^{-i\theta/2}
\end{array}\right),
\end{equation}
where $\theta$ is the strong CP theta-angle.

If we consider the up-type quarks for a moment, we have that 
 $\boldsymbol{\tau}$ and $\boldsymbol{\nu}$ lie in the 
same plane as $\mathbf{v}_c$ and $\mathbf{v}_u$ at $\mu=m_t$, and so the Darboux 
triad must be related to the state vector triad by a rotation 
in this plane, 

\begin{equation}
\Omega_U=\left(\begin{array}{ccc}
1&0&0\\
0&\cos\omega_U&-\sin\omega_U\\
0&\sin\omega_U&\cos\omega_U\end{array}\right).
\end{equation}
Here $\omega_U$ is the angle between $\boldsymbol{\tau}$ and 
$\mathbf{v}_c$.  Following \cite{Bordes:2010nb} we find that the CKM 
matrix is given by

\begin{equation}
V_{\textrm{CKM}} = (\Omega^{-1}_UP_0\Omega_U)V(\Omega^{-1}_DP^{-1}_0\Omega_D)
\end{equation}
where $V$ is the unitary matrix in \eqref{up_down}.  
The mass formula 
\eqref{masses} is unaltered by this chiral transformation.  For 
given $V$, $\omega_U$ and $\omega_D$ 
the experimental value of the Jarlskog invariant can then be used to 
fix $\theta$.

Assuming that the phases in $\balpha$ do not change with
scale
is equivalent to assuming that all CP violating effects in the CKM
matrix 
come from the strong theta-angle term in the manner set out above.
Since the masses and mixing angles depend only on the inner products
of $\balpha$ with the state vectors the phases will cancel out in this
case.  This means we can take these vectors to be real without loss of
generality. 
For simplicity we shall do so in our fit below and $V$ then becomes an
orthogonal matrix.  

We could consider the leptonic sector analogously.  However, it is not known whether neutrino oscillations violate CP symmetry.  It is also not known if neutrinos are Majorana particles, which would cause extra phases to enter the MNS matrix.  As experiments so far have nothing to say about these points, and very little is known about neutrino masses, we do not consider the neutrinos in this work.

\subsection{Fitting $\balpha(\mu)$ to data}

The rotating mass matrix scheme has $\boldsymbol{\alpha}$ as a 
fundamental object, and the state vectors are derived from this.
To test this hypothesis we shall start from real orthonormal state
vectors and use experimental data to find a consistent trajectory of 
$\balpha$.  We first choose the up type quarks to have state vectors:

\begin{align}
\mathbf{v}_{u}& = (1,\ 0,\ 0)^\dagger,\\
\mathbf{v}_{c}& = (0,\ 1,\ 0)^\dagger,\\
\mathbf{v}_{t}& = (0,\ 0,\ 1)^\dagger,
\end{align}
which we are free to do.  Once we have chosen a $V$ we can define the down type quark state vectors, $\mathbf{v}_{d,s,b}=V\mathbf{v}_{u,c,t}$.  There are no explicit physical constraints on the matrix $V$ so we are free to choose any orthogonal matrix.

For the top quark $\balpha(\mu=m_t) = \mathbf{v}_{t}$, and similarly $\balpha(\mu=m_b) = \mathbf{v}_{b}$.  The leakage mechanism then fixes $\balpha_{c}$ and $\balpha_{s}$ in terms of the quark masses:

\begin{align}
\balpha_{c}& = \sqrt{m_{c}/m_{t}}\  \mathbf{v}_{c} +  \sqrt{1 - m_{c}/m_{t}}\  \mathbf{v}_{t},\\
\balpha_{s} &= \sqrt{m_{s}/m_{b}}\  \mathbf{v}_{s} +  \sqrt{1 - m_{s}/m_{b}}\  \mathbf{v}_{b}.
\end{align}
The two vectors $\mathbf{v}_{t}$ and $\mathbf{v}_{c}$ define a plane.  All that the mass ratio $m_u/m_t$ tells us about $\balpha_{u}$ is the angle which it makes with this plane.  We can thus restrict $\balpha_{u}$, and similarly $\balpha_{d}$, to lie somewhere on a line, parametrised by $t \in [0,\ 2\pi)$:

\begin{align}
\balpha_{u}& = \sqrt{m_{u}/m_{t}}\ \mathbf{v}_u + \sqrt{1 - m_{u}/m_{t}}\  \mathbf{v}_c\cos (t) + \sqrt{1 - m_{u}/m_{t}}\ \mathbf{v}_t \sin (t),\\
\balpha_{d}& = \sqrt{m_{d}/m_{b}}\ \mathbf{v}_d + \sqrt{1 - m_{d}/m_{b}}\  \mathbf{v}_s\cos (t) + \sqrt{1 - m_{d}/m_{b}}\ \mathbf{v}_b \sin (t),
\end{align}
where we will choose the signs of the square roots so that we obtain a right handed triad.  The restrictions on $\balpha_{i}$ for the charged leptons were found in an analogous way, replacing $\left( u, c, t\right)$ with $\left( e, \mu, \tau\right)$.

\begin{figure}
\begin{center}
\begin{tabular}{cc}
\hspace{-0cm}
\includegraphics[height=6cm]{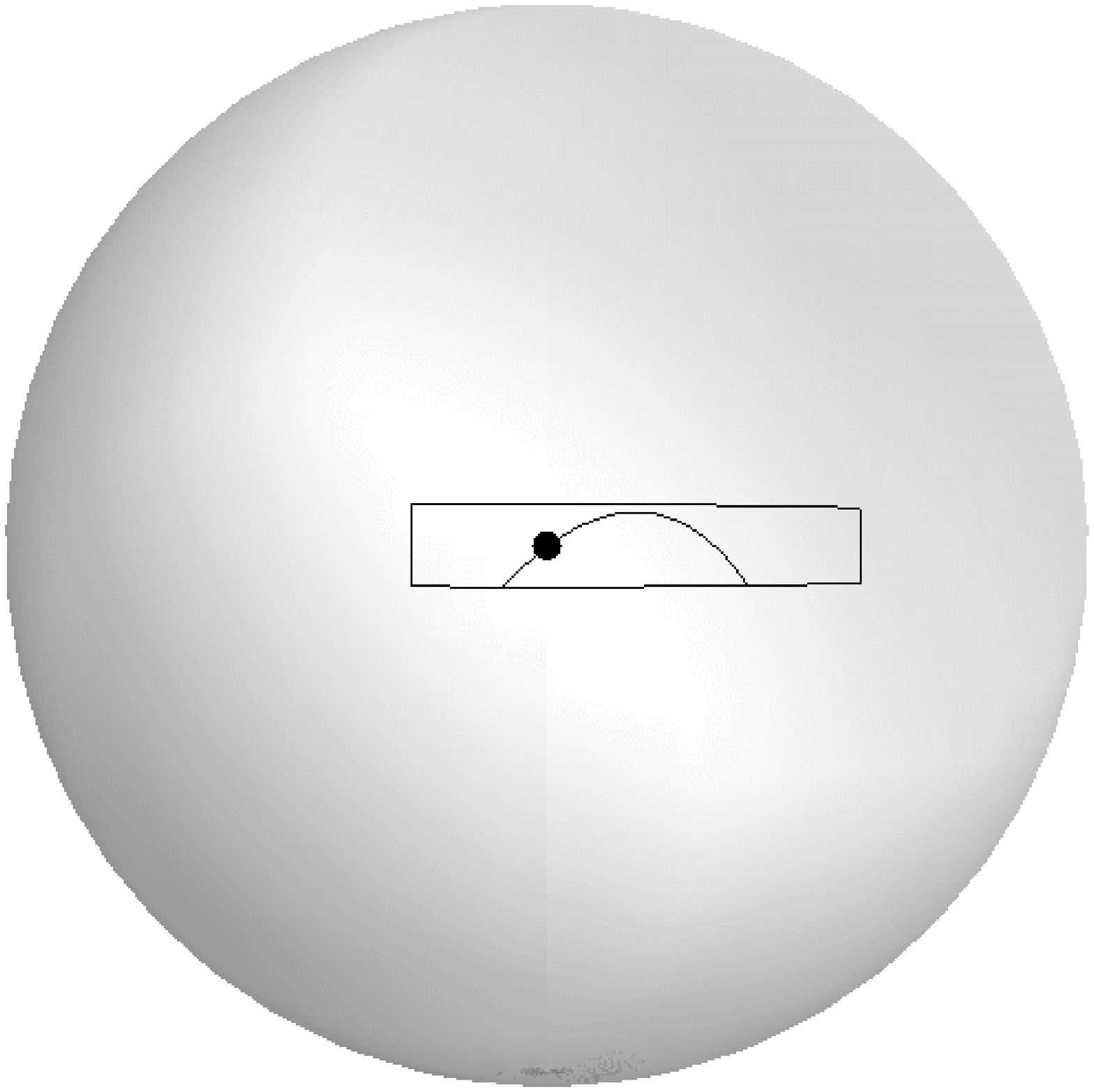}&
\hspace{-1cm}
\includegraphics[height=6cm]{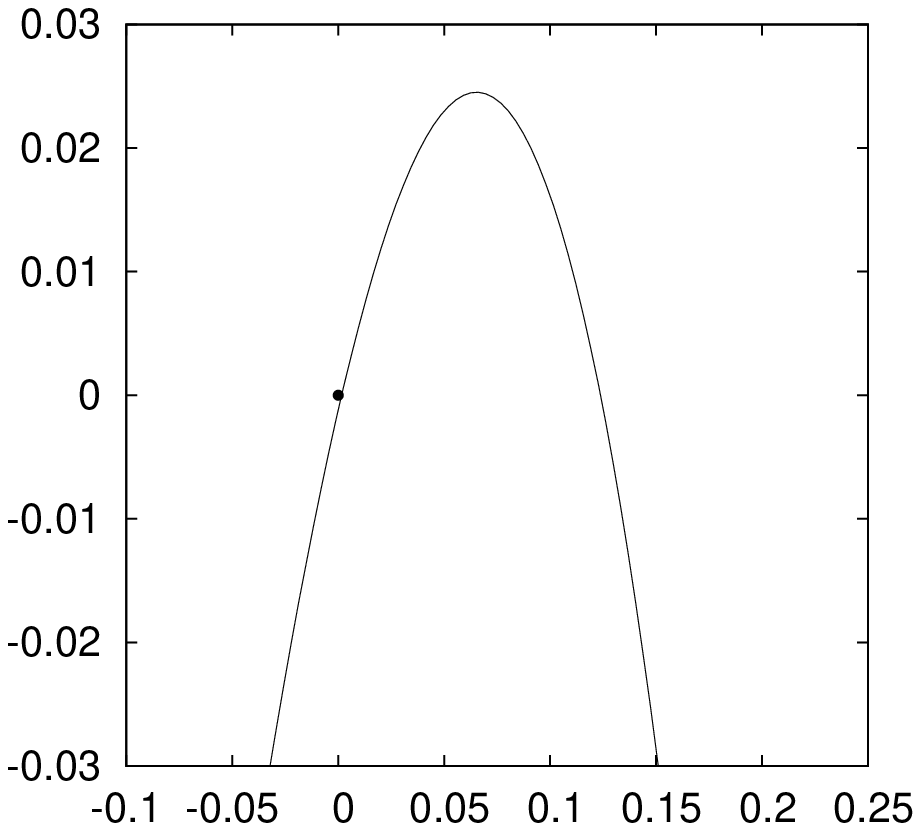}
\end{tabular}
\caption{The state vectors and $\balpha$ range over a unit sphere in $\mathbb{R}^3$.  These are conveniently represented as points on the plane under stereographic projection.  The point shown lies at the south pole of the projection.  The box on the sphere is the boundary of the plot on the right, and the curve shows the best fit line found.}
\label{fig:SP}
\end{center}
\end{figure}

Before we give any results we will take a moment to discuss their presentation.  The state vectors and $\balpha$ are in $\mathbb{R}^3$ and of unit length so take values on the surface of a unit sphere.  We will represent their positions by stereographically projecting onto $\mathbb{R}^2$.  It turns out that $\balpha$ does not need to rotate very far from $\mu=m_t$ to $\mu=m_e$ so most of the action happens in a small area on the sphere.  We have chosen the south pole of the projection to be at the position of $\mathbf{v}_\tau$.  This means that there will not be much distortion introduced by the stereographic projection; geodesics in this region on the sphere will be almost straight lines on the plane.  Figure \ref{fig:SP} shows the unit sphere with the region we will be interested in enclosed in a box.  The curve within the box shows the best fit line we find and the point shows the south pole of the projection, $\mathbf{v}_{\tau}$.  The projection itself is shown on the right.  The metric on the sphere is given by

\begin{equation}
ds^2=\frac{4}{(1+u^2+v^2)^2}(du^2+dv^2)
\end{equation}
for coordinates on the plane $u$ and $v$.  Over the boundary box in
figure \ref{fig:SP} the metric ranges from $4(du^2+dv^2)$ to $3.76(du^2+dv^2)$; there is a maximum distortion of a length in the stereographic projection of $3\%$.

Though we assume that the trajectory of $\balpha$ is universal, there is no physical constraint on the relation between the quark and lepton sectors.  We thus have the freedom to match these two sectors to give the smoothest trajectory.  As mentioned previously we also have a freedom in choosing the matrix $V$.  To find the best matching we ranged over, and then applied simplex optimisation at good regions in, the parameter space of quark-lepton sector matching matrices and quark mixing matrices, $V$.  The positions of $\balpha_{t,\, c,\, b,\, s,\, \tau,\, \mu}$ were found and projected onto the plane.  We then fit a cubic line to these points on the plane using a non-linear least squares algorithm.  Previous work \cite{Bordes:2002nh} has shown that the cumulative arc length between $\balpha_x$ can be modelled by an exponential function at high scales.  Accordingly, we then fit an exponential function to the cumulative arc length, excluding the strange quark and the electron.  The strange quark was excluded since the interpretation of its intermediate mass is somewhat uncertain in this scheme.  The electron was excluded firstly since we are only interested in the relationship at higher scales and secondly since we do not have good restrictions on the cumulative arc length, as $\balpha_e$ can only be constrained to lie on a line.  Finally we use the Jarlskog invariant to fix the value of $\theta$ and so determine the magnitudes of the elements of the CKM matrix.  We chose the quark-lepton sector matching and the matrix $V$ which produced absolute values of the CKM elements close to the experimentally measured magnitudes whilst maximising the R-Squared value of both the cubic and the exponential fit.  The R-Squared value of a fit is given by

\begin{equation}
R^2\equiv1-\frac{\sum_i^N\left(y_i-f_i\right)^2}{\sum_i^N\left(y_i-\bar{y}\right)^2}
\end{equation}
where $y_i$ are the $y$ coordinates of $N$ data points, $f_i$ are the $y$ coordinates of the best fit line at the same $x$ coordinates and $\bar{y}=\frac{1}{N}\sum_i^N y_i$.

\begin{figure}
\begin{center}
\begin{tabular}{cc}
\multicolumn{2}{c}{\hspace{-2cm} \includegraphics[width=13cm]{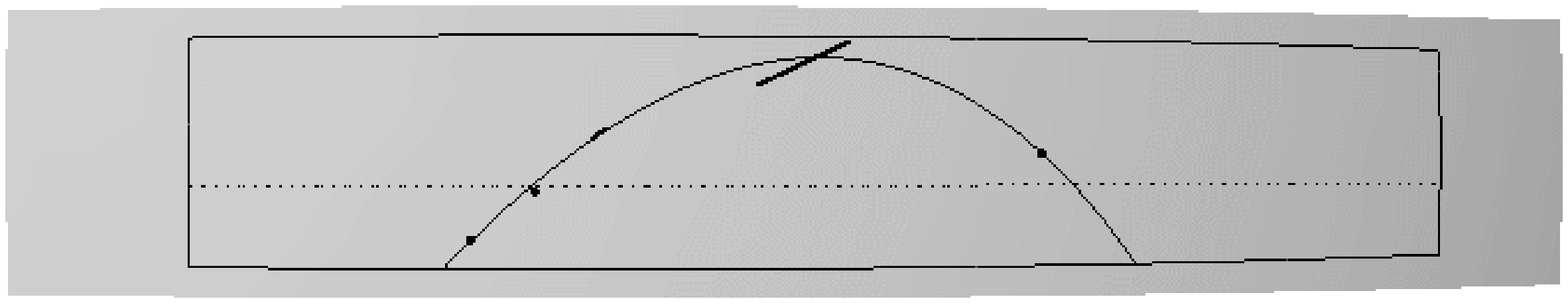} \vspace{1cm}}\\
\hspace{-1.5cm}
\includegraphics[width=8cm]{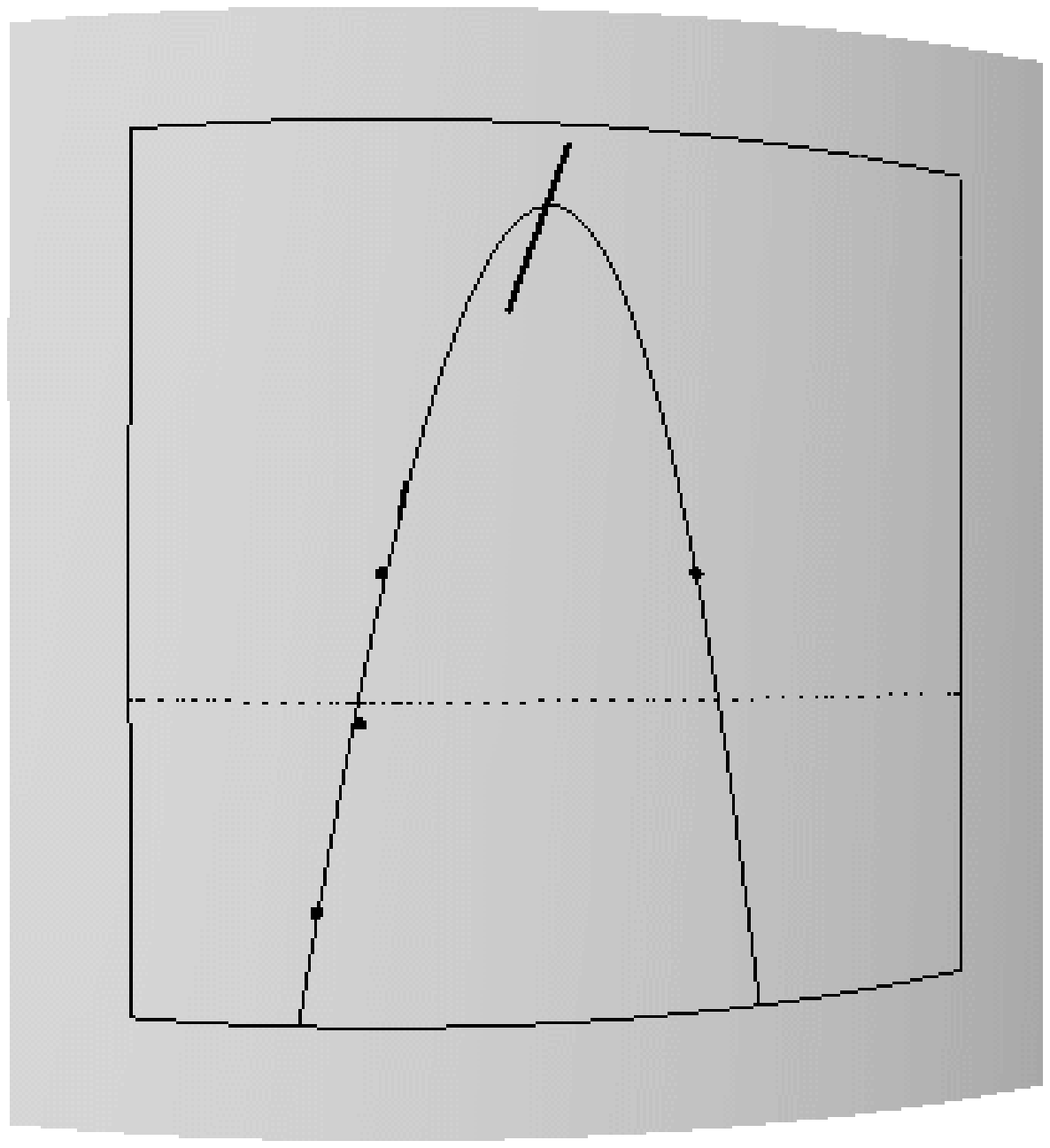}&
\hspace{-2cm}
\begin{overpic}[width=11cm]{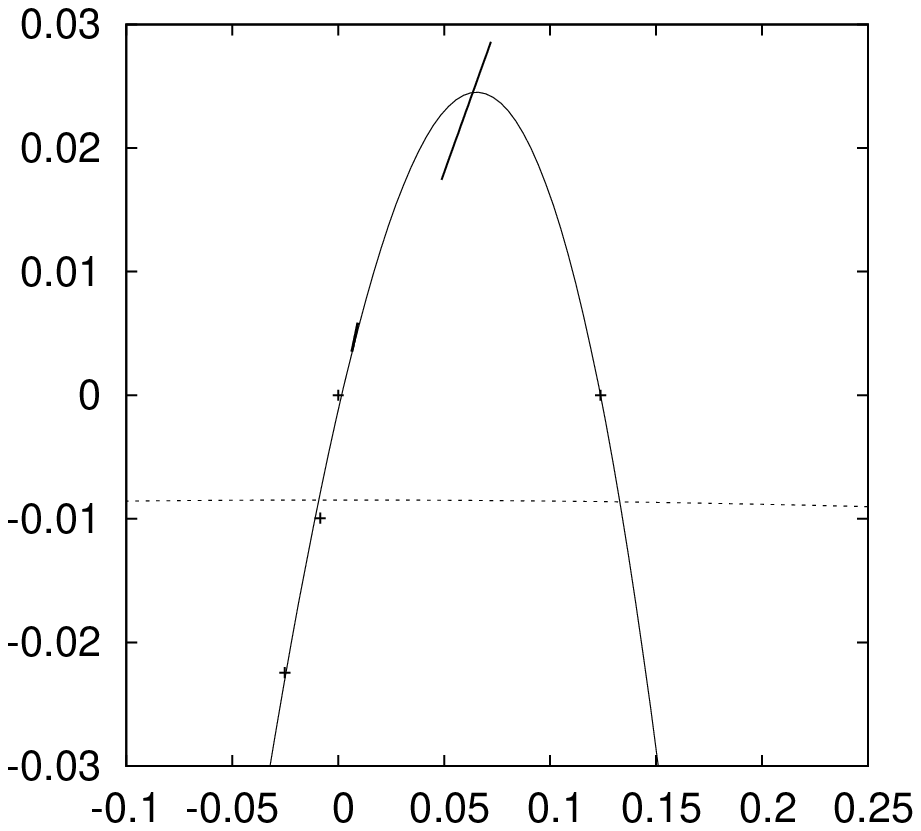}
\put(32,13){$\balpha_t$}
\put(42,24){$\balpha_b$}
\put(36,36){$\balpha_\tau$}
\put(38,41){$\balpha_c$}
\put(52,53){$\balpha_s$}
\put(65,36){$\balpha_\mu$}
\put(86,27){$\balpha_e$}
\end{overpic}
\end{tabular}
\caption{The positions of $\balpha$ at various scales determined partly by experimental constraints and partly by our choices as described in the text.  The position of $\balpha_e$ is constrained to lie somewhere on the dotted line.  The cubic best fit line is shown by the solid line.  The top shows these on the sphere, left shows them on an ellipsoid with the axes stretched to match those of the stereographic projection while the right shows the stereographic projection with the positions of $\balpha_x$ indicated.}
\label{fig:Best_Fit}
\end{center}
\end{figure}

Figure \ref{fig:Best_Fit} shows the cubic best fit line along with the positions of $\balpha_x$.  The experimental errors in the masses of the quarks lead to an uncertainty in the position of $\balpha_c$ and $\balpha_s$.  The 1 $\sigma$ errors in the quark masses restrict $\balpha_c$ and $\balpha_s$ to lie on the lines shown.  The cubic best fit line is

\begin{equation}
0.75\, x-4.83\, x^2-9.19\, x^3.
\end{equation}

The cumulative arc length between $\balpha_x$'s was found after mapping the cubic best fit line onto the sphere through an inverse stereographic projection.

\begin{figure}
\begin{center}
\begin{overpic}[width=10cm]{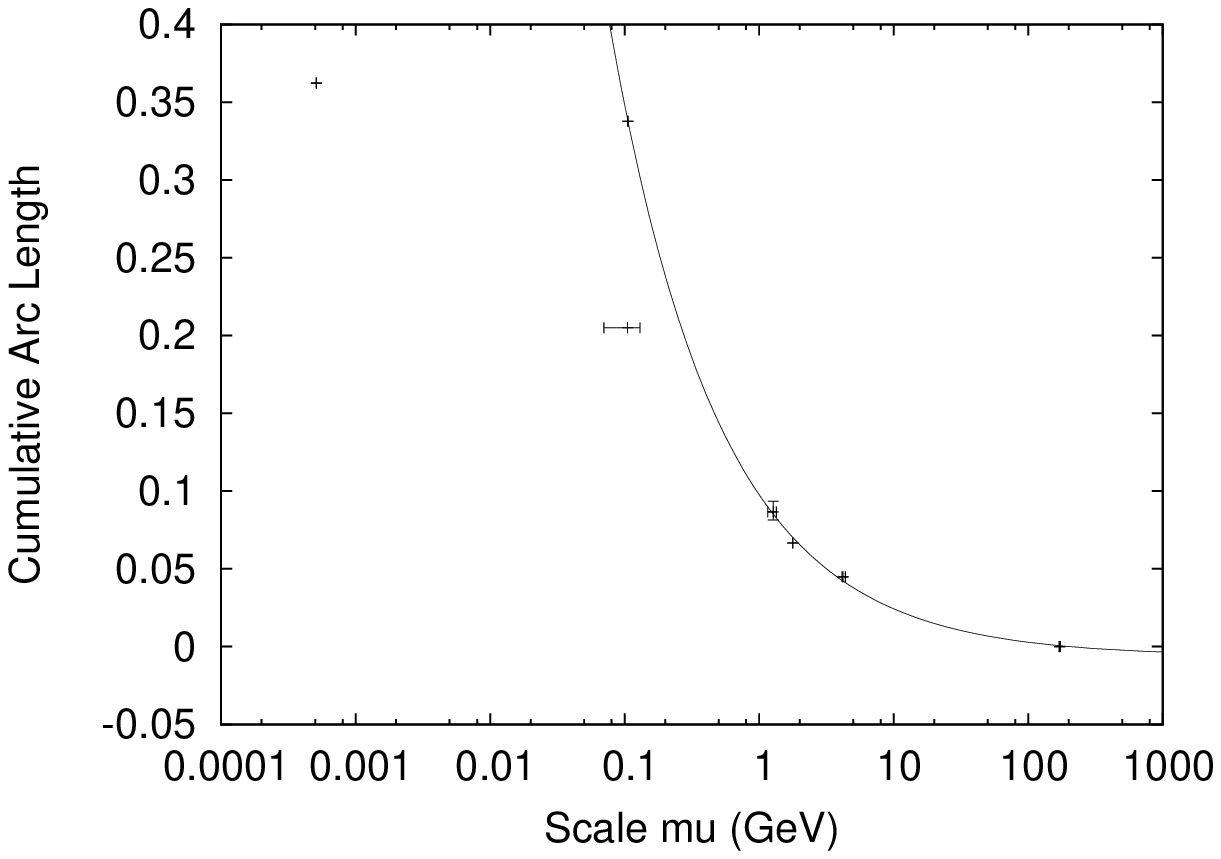}
\put(83,18){$\balpha_t$}
\put(64,19){$\balpha_b$}
\put(60,22){$\balpha_{\tau}$}
\put(57,26){$\balpha_c$}
\put(44,41){$\balpha_s$}
\put(54,58){$\balpha_{\mu}$}
\put(29,61){$\balpha_e$}
\end{overpic}
\caption{The cumulative arc length along the best fit line measured from $\balpha_t$ is well approximated by an exponential curve for all but $\balpha_s$ and $\balpha_e$.  A great circle gives an arc length of $2 \pi$ in these units. }
\label{fig:Arc_Length_Line}
\end{center}
\end{figure}

Figure \ref{fig:Arc_Length_Line} shows that the arc lengths between $\balpha_t$, $\balpha_b$, $\balpha_{\tau}$, $\balpha_c$ and $\balpha_{\mu}$ are well fitted by the exponential curve
\begin{equation}
0.104\exp\left(-1.228\log_{10}\left(\mu\right)\right)-0.0061
\end{equation}
for $\mu$ in GeV.  This is in good agreement with the results from the 
planar approximation found in \cite{Bordes:2002nh}.  The point 
$\balpha_s$ does not fit on the exponential curve.  This is expected,
as the light mass of the $s$ quark means that the definition of mass
we use is not entirely correct for it. 
The arc length to $\balpha_e$ is estimated by the arc length to $\balpha_e$ line.  This is just an estimate since $\balpha_e$ may sit anywhere on this line.  It is however clear that $\balpha_e$ is unlikely to sit on the exponential curve found.  Various models which employ the rotating mass matrix mechanism suggest a $\tanh(\mu)$ like behaviour, with fixed points for the rotation at $\mu=0$ and $\mu=\infty$, so it is not surprising that the exponential fit matches the data well at higher scales but not at lower scales.  For the later work on Higgs decay we only need to model the behaviour at high scales.

\begin{figure}
\begin{center}
\begin{tabular}{cc}
\hspace{-2.6cm}
\begin{overpic}[width=10cm]{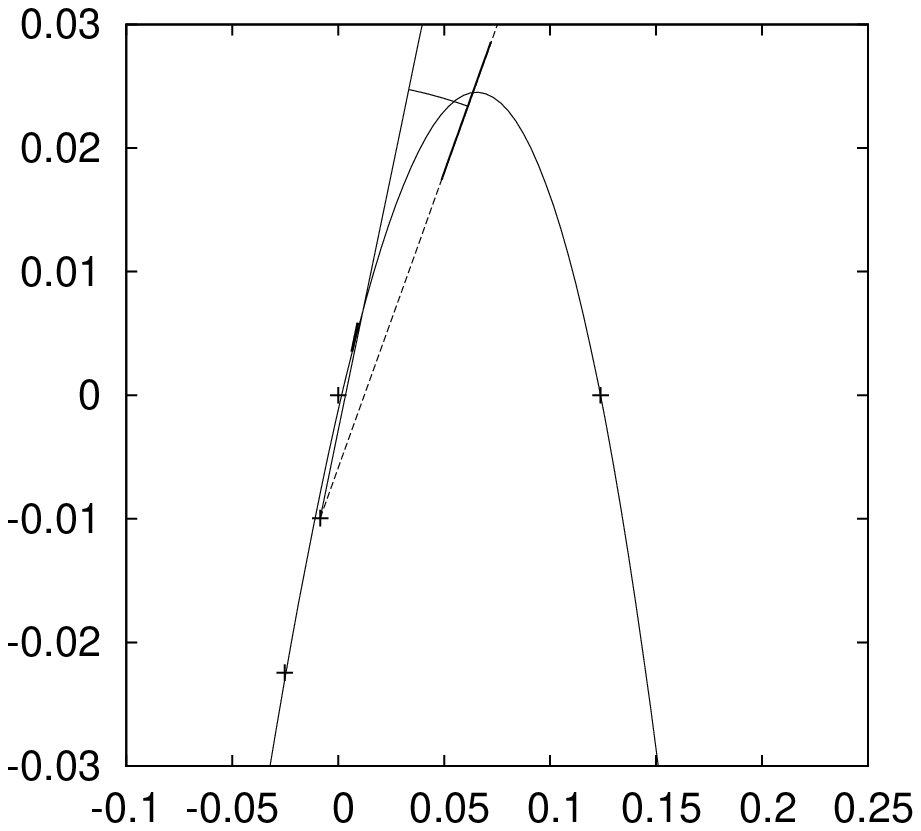}
\put(48.7,62.5){$\omega_D$}
\end{overpic}&
\hspace{-3cm}
\begin{overpic}[width=10cm]{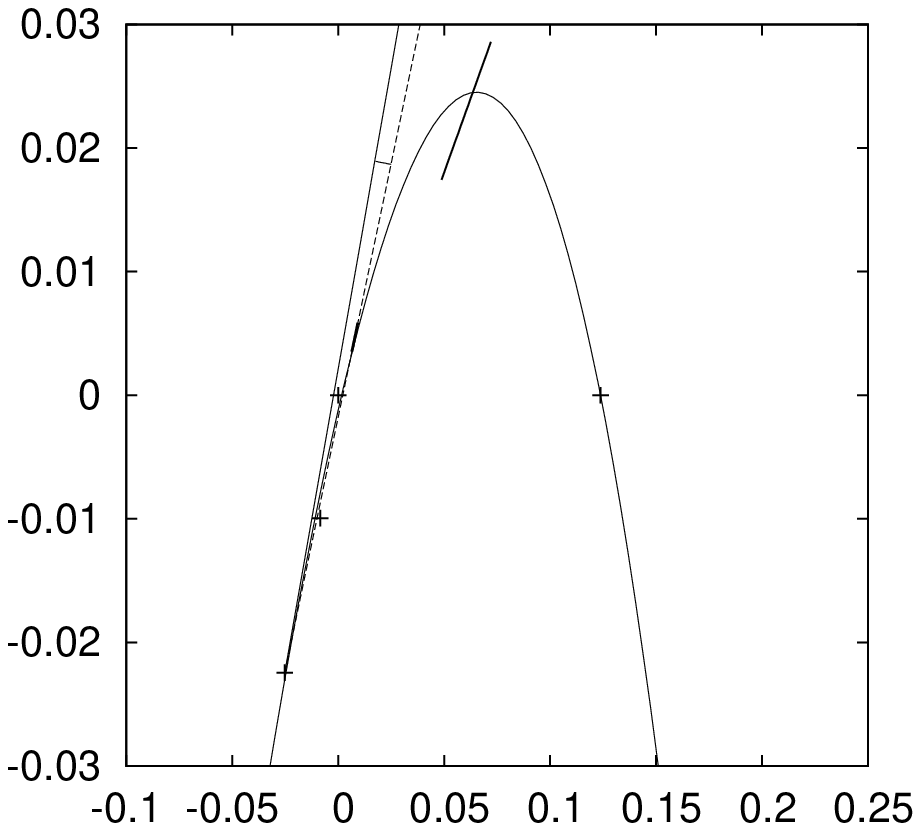}
\put(42.2,56.2){$\searrow$}
\put(37.8,59){$\omega_U$}
\end{overpic}
\end{tabular}
\caption{An illustration of the angle $\omega_D$ ($\omega_U$).  Here the state vector triad and the Darboux triad have their origins at $\balpha_{b}$ ($\balpha_{t}$).  The solid line shows $\boldsymbol{\tau}(\mu)\ ||\ \dot{\balpha}(\mu)$ and the dashed line shows $\mathbf{v}_{s}$  ($\mathbf{v}_{c}$).  It should be remembered that the axes are not of equal scale and that this is a stereographic projection of the vectors so the angles cannot be directly read off.}
\label{fig:Omega}
\end{center}
\end{figure}
For this trajectory we find $\omega_U = 0.09$ radians and $\omega_D = 0.25$ radians.  Fitting a Jarlskog invariant of $J=3.05\times10^{-5}$ gives a strong CP angle of $1.45$ radians.  These results are in line with estimates in \cite{Bordes:2010nb}.  The absolute values of the CKM matrix obtained are:
\begin{equation}
\left(\begin{array}{ccc}
0.97430&0.2252&0.00357\\
0.2251&0.97345&0.0415\\
0.00879&0.0407&0.999134\\
\end{array}\right),
\end{equation}
which can be compared with the experimental values \cite{Amsler:2008zzb}:
\begin{equation}
\left( \begin{array}{ccc}
0.97419\pm0.00022&0.2257\pm0.0010&0.00359\pm0.00016\\
0.2256\pm0.0010&0.97334\pm0.00023&0.0415^{+0.0010}_{-0.0011}\\
0.00874^{+0.00026}_{-0.00037}&0.0407\pm0.0010&0.999133^{+0.000044}_{-0.000043}
\end{array} \right).
\end{equation}
We find the unitarity angles, defined and measured \cite{Amsler:2008zzb} as
\begin{eqnarray}
\alpha=\arg\left(-\frac{V_{td}V_{tb}^*}{V_{ud}V_{ub}^*}\right)&=&(88^{+6}_{-5})^\circ,\\
\beta=\arg\left(-\frac{V_{cd}V_{cb}^*}{V_{td}V_{tb}^*}\right)&=&\frac{1}{2}\sin^{-1}(0.681\pm0.025),\\
\gamma=\arg\left(-\frac{V_{ud}V_{ub}^*}{V_{cd}V_{cb}^*}\right)&=&(77^{+30}_{-32})^\circ,
\end{eqnarray}
to be $\alpha=88^\circ$, $\sin(2\beta)=0.691$ and $\gamma=70^\circ$.

\section{Higgs Decay Branching Ratios}

Now that we have found a trajectory of $\balpha$ which agrees well with experimental constraints we can use it to ask what the rotating mass matrix hypothesis may say about Higgs decay.  Though we have so far ignored the lightest quarks note that the up and down triads are fixed at $\mu=m_c$ and $\mu=m_s$ respectively.  We can make predictions for the branching ratios of modes involving the up and down quarks since we shall see that they depend only on the quark state vectors and the trajectory of $\balpha$ for $\mu \geq m_H/2$.  We can similarly make predictions for modes involving the electron.  Previously \cite{Bordes:2009wi} the following Yukawa couplings have been suggested, e.g., for the up type quarks:
\begin{equation}
\mathcal{A}_{YK}=\rho_U\balpha\bar{\psi}_L\phi\psi_R\balpha^\dagger + h.c.
\end{equation}
Choosing the gauge in which $\phi$, the Higgs doublet, is real and
points in the up direction and expanding the remaining real component
about its minimum value $\zeta_W$, $\phi^1_R = \zeta_W + H$, we obtain 
the zeroth order mass matrix, equation \eqref{rank_one_mass_matrix}, with
$m_U=\rho_U\zeta_W$, and the first order coupling matrix of the Higgs 
boson to the quarks as
\begin{equation}
Y=\rho_U\balpha\balpha^\dagger.
\end{equation}
Since this Yukawa coupling matrix depends on scale, $\mu$, we need to
take care in defining the details of Higgs decay.  In constraining the
trajectory of $\balpha$ we were considering fermion masses, whereas
now we are considering Higgs decay.  These are clearly related
processes but in integrating the assumed underlying RGE the constant
of integration may differ.  It turns out that 
the correct calibration is a factor of 2 change in scale, i.e.,
$\balpha_H=\balpha(\mu=m_H/2)$ \cite{Bordes:2009wi}.  We take the 
scale of Higgs decay to be that of the Higgs mass, as usual.  We can
now 
use the best fit line to find the Higgs state tensor, $\balpha\balpha^\dagger$, as a function of Higgs mass.

We then get the coupling for Higgs decaying into $x\bar{y}$ as

\begin{equation}
A(H\rightarrow x\bar{y})=\rho_T | \mathbf{v}_x\cdot\balpha_H || \mathbf{v}_y\cdot\balpha_H |.
\end{equation}
We can ignore any kinematic factors by considering ratios of branching ratios:

\begin{equation}
\label{branching_ratios}
\frac{\Gamma(H\rightarrow x\bar{y})}{\Gamma(H\rightarrow b\bar{b})}=
\frac{\rho^2_T}{\rho^2_D}\frac{| \mathbf{v}_x\cdot\balpha_H|^2| \mathbf{v}_y\cdot\balpha_H|^2}{| \mathbf{v}_b\cdot\balpha_H|^4}.
\end{equation}

LEP found a lower limit on the Higgs mass of 114.4 GeV at 95\% confidence level \cite{Amsler:2008zzb}.  Various upper bounds have been given for the standard model Higgs boson mass. Here we plot up to $m_H=260$ GeV.

\begin{figure}
\begin{center}
\begin{tabular}{cc}\hspace{-2.5cm}
\begin{overpic}[width=9cm]{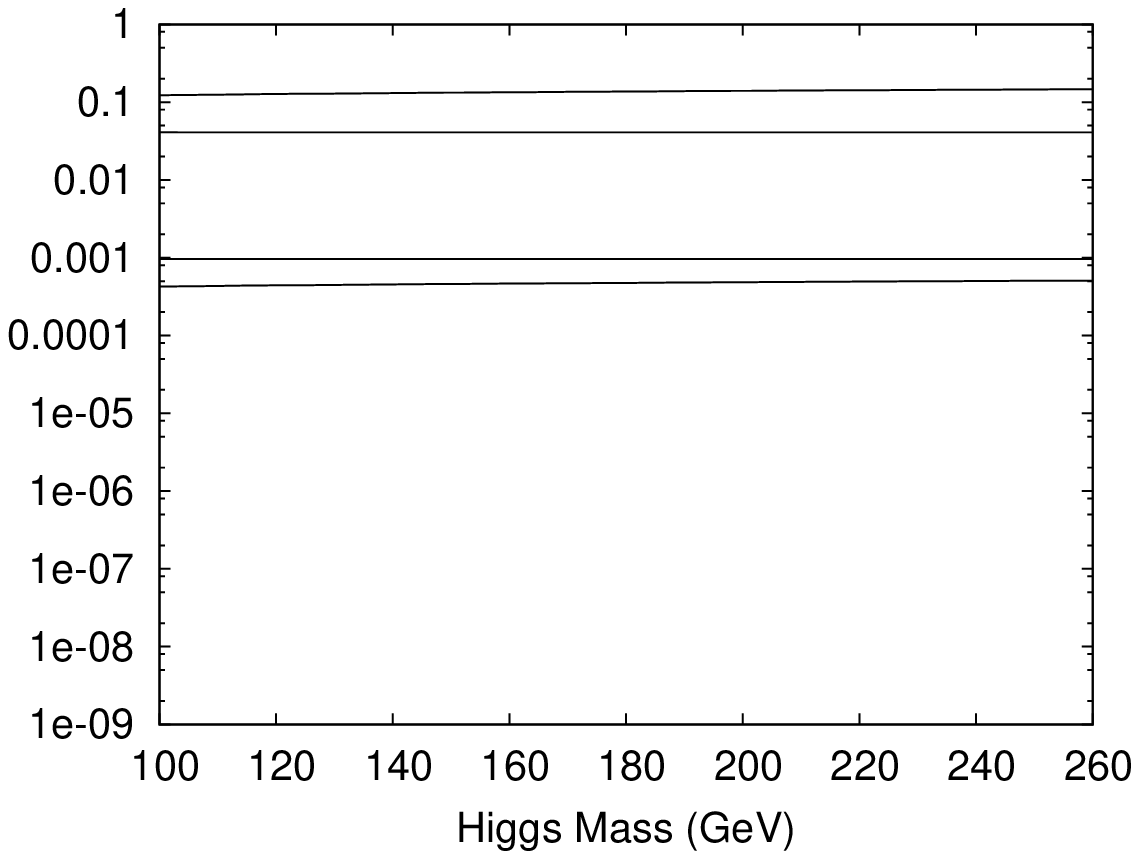}
\put(68,62){$\tau^-\tau^+$}
\put(68,53.5){$c\bar{c}$}
\put(78,49){$s\bar{s}$}
\put(78,41){$\mu^-\mu^+$}
\end{overpic}
&\hspace{-1cm}
\begin{overpic}[width=9cm]{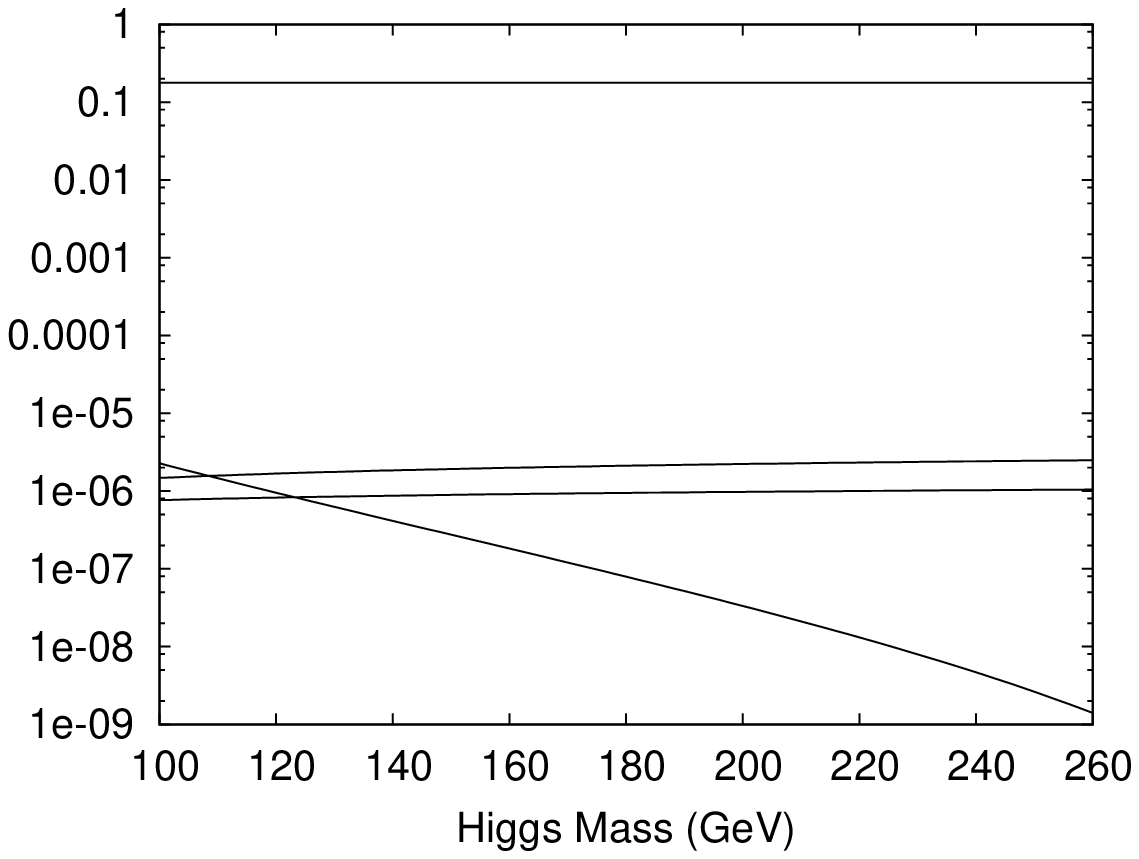}
\put(75,57){$\tau^-\tau^+$}
\put(50,18){$c\bar{c}$}
\put(75,33){$s\bar{s}$}
\put(75,24){$\mu^-\mu^+$}
\end{overpic}
\end{tabular}
\caption{$\Gamma(H\rightarrow x\bar{x})/\Gamma(H\rightarrow b\bar{b})$ for various final state particles as predicted by the standard model (left) and the rotating mass matrix hypothesis (right).}
\label{fig:BRSM}
\end{center}
\end{figure}
Since
now we have explicit formulae for the best fit trajectory of $\balpha$
and the relation between arc length and scale, at high energies, it 
becomes a simple matter to find the ratios of branching ratios, given by \eqref{branching_ratios}, for a range of Higgs masses.  Figure \ref{fig:BRSM} shows the standard model predictions for Higgs decay along with the predictions from the rotating mass matrix hypothesis.  The standard model predictions were found using HDECAY \cite{Spira:2009}.  We can see that the $c\bar{c}$ decay mode is heavily suppressed, in accordance with the estimate in \cite{Bordes:2009wi}.  This mode is suppressed since near $\mu=2\, m_t$ the eigenvector $\balpha$ is almost orthogonal to $\mathbf{v}_c$, reducing the branching ratio to $c\bar{c}$.  The suppression is to such a degree over the whole range of Higgs masses that detection should not require large statistics.  The $s\bar{s}$ and $\mu^-\mu^+$ modes are also suppressed, though to a smaller degree.

\begin{figure}
\begin{center}
\begin{tabular}{cc}\hspace{-2.5cm}
\begin{overpic}[width=9cm]{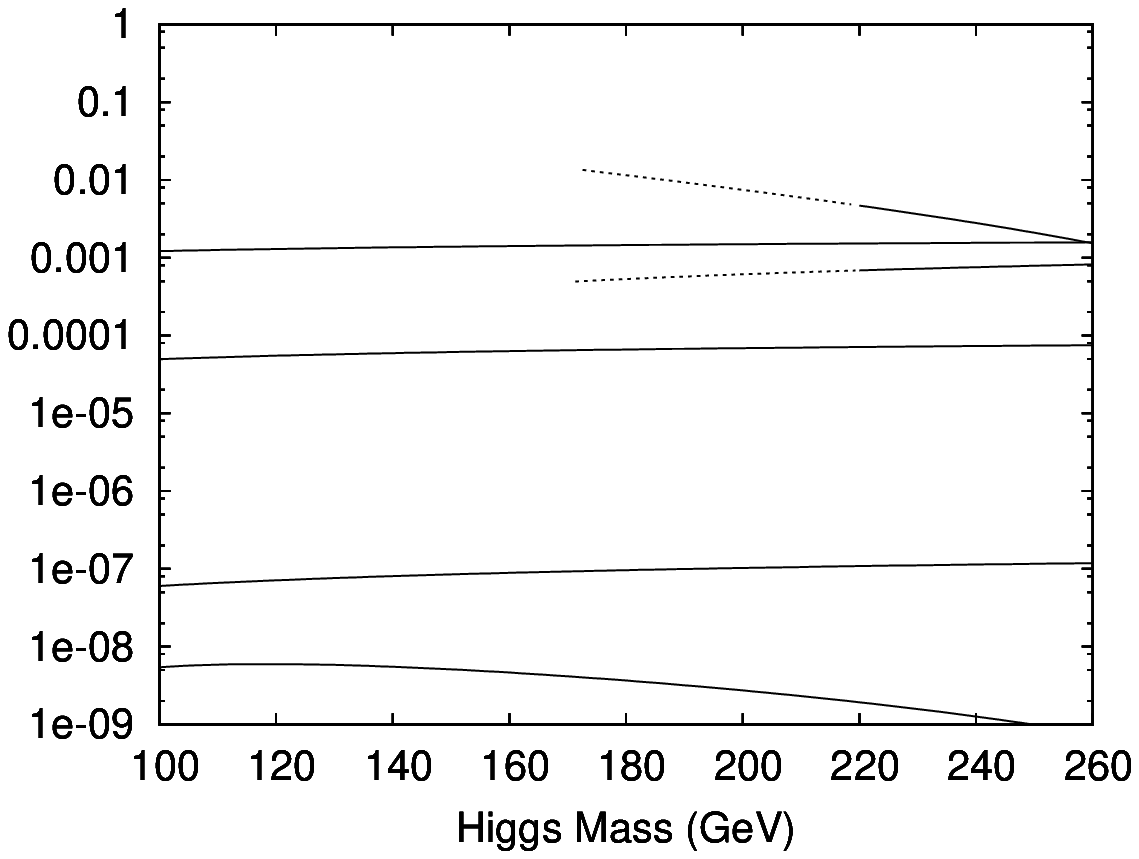}
\put(30,49){$b\bar{s}$}
\put(30,41){$b\bar{d}$}
\put(68,24){$s\bar{d}$}
\put(68,14){$c\bar{u}$}
\put(48,53){$t\bar{c}$}
\put(47,44){$t\bar{u}$}
\end{overpic}
&\hspace{-1cm}
\begin{overpic}[width=9cm]{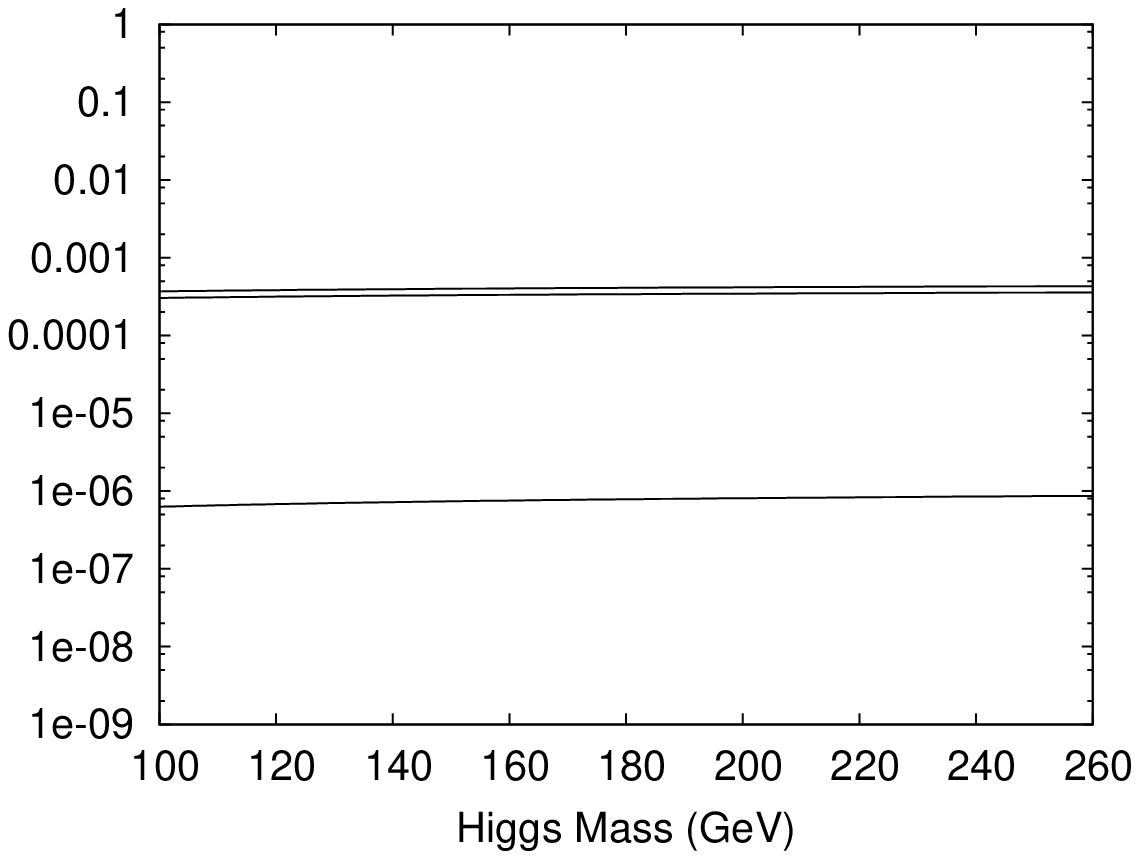}
\put(75,47){$\tau^-\mu^+$}
\put(75,40){$\tau^-e^+$}
\put(75,30){$\mu^-e^+$}
\end{overpic}
\end{tabular}
\caption{$\Gamma(H\rightarrow x\bar{y})/\Gamma(H\rightarrow b\bar{b})$ for various flavour violating decays as predicted by the rotating mass matrix hypothesis.  Note that $\Gamma(H\rightarrow y\bar{x})=\Gamma(H\rightarrow x\bar{y})$.  Below around $220$ GeV, indicated by the dotted lines, threshold effects will influence the $t\bar{c}$ and $t\bar{u}$ decay modes.}
\label{fig:BRFV}
\end{center}
\end{figure}

Although we propose no dynamical mechanism the rotating mass matrix picture generically predicts flavour violating decays, shown in figure \ref{fig:BRFV}.  If the Higgs mass is large then $t\bar{c}$ and $t\bar{u}$ decay modes are possible.  To be in line with the rest of the estimates, where the decays are well above threshold, we neglect threshold effects for these modes.  Above $220$ GeV we believe that these effects will be negligible and that the estimates shown for the branching ratios are realistic.  The branching ratio for $H\rightarrow \tau^-\mu^+$ is almost three orders of magnitude higher than that for $H\rightarrow \mu^-\mu^+$.  This is in stark contrast to the standard model predictions where $H\rightarrow \tau^-\mu^+$ cannot occur at tree level and so has a small branching ratio.  These modes may have a cleaner signal than $H\rightarrow c\bar{c}$, making it easier to detect.  Flavour violating effects at the levels predicted here have been shown in \cite{Bordes:2009wi} to be consistent with existing experimental bounds on flavour violation.

\section{Conclusions}

We have found that the rotating mass matrix hypothesis can accommodate the CKM matrix with a CP violating phase and, with a few caveats, match the experimental data.  The $\balpha_x$'s can be placed on a smooth trajectory so that they give the correct masses, for all but the light quarks, and give a good CKM matrix. Fitting the Jarlskog invariant gives a theta-angle of 1.45 radians.  

We have found that for the heavy quarks and the charged leptons a simple exponential fit can model the cumulative arc length for scales above $m_\mu$, though it is unlikely to be a realistic fit below this scale.  This behaviour has been captured by both a phenomenological model (DSM \cite{Bordes:1997ft}) and a field theory (MBSM \cite{Chan:2006nf}), which have rotating mass matrices and predict a $\tanh(\mu)$ like behaviour with fixed points in the rotation at $\mu = 0$ and $\mu=\infty$.

Since we do not yet know whether neutrino oscillations violate CP symmetry, nor whether they are Majorana particles, the neutrinos place no constraints on $\balpha(\mu)$ in the scheme as envisaged here.

The cubic best fit line allows us to predict branching ratios for a range of Higgs masses.  These give $H\rightarrow c\bar{c}$ suppression, in line with a previous planar approximation \cite{Bordes:2009wi}, along with $\mu^-\mu^+$ and $s\bar{s}$ suppression.  We find a notable braching ratio to $\tau^-\mu^+$ and give limits for other flavour violating decays.

The authors would like to acknowledge the contributions of Chan
Hong-Mo and Jos\'{e} Bordes both in comments on the present work and
in continual communications.  One of the authors (MJB) would also like to acknowledge financial support from EPSRC.

\end{document}